\documentclass[10pt,twoside,BCOR7mm,DIV15,headinclude,footexclude,cleardoubleempty,idxtotoc]{scrartcl}

\usepackage[english]{babel}
\usepackage{graphicx}
\usepackage{hyperref}
\usepackage{scrpage2}
\usepackage{hyperref}
\usepackage{ifthen}

\makeatletter
\renewcommand{\@biblabel}[1]{}
\renewcommand{\@cite}[2]{%
{#1\ifthenelse{\boolean{@tempswa}}{,#2}{}}}
\makeatother

\hypersetup{breaklinks=true
,colorlinks=true,linkcolor=black,urlcolor=blue
,citecolor=black}

\ofoot{\thepage}
\ifoot{}

\setheadsepline{1pt}

\setkomafont{pagehead}{\normalfont\sffamily}
\setkomafont{pagenumber}{\normalfont\rmfamily}

\usepackage{booktabs}
\usepackage{amsmath}
\usepackage{amssymb}
\usepackage{multicol}
\usepackage{float}

\makeatletter
\newcommand{\listofcontributions}{\@starttoc{con}}

\newcommand{\l@contribution} {\@dottedtocline{1}{1.5em}{2.3em}}
\makeatother

\newenvironment{contribution}{
\setcounter{section}{0}
\setcounter{figure}{0}
\setcounter{table}{0}
\begin{flushleft}
{\em Clumping in Hot Star Winds \\
W.-R.\ Hamann, A.\ Feldmeier \& L.\ Oskinova, eds.\\
Potsdam: Univ.-Verl., 2007 \\
URN: http://nbn-resolving.de/urn:nbn:de:kobv:517-opus-13981
} 
\end{flushleft}
}{
\newpage
\lehead{}
\rohead{}
}

\begin{document}

\setlength{\baselineskip}{2.5ex}

\begin{contribution}

\lehead{Urbaneja, Kudritzki \& Puls}

\rohead{Clumping in the winds of O-type CSPNs}

\begin{center}
{\LARGE \bf Clumping in the winds of O-type CSPNs}\\
\medskip

{\it\bf M. A. Urbaneja$^1$, R.-P.\ Kudritzki$^1$ \& J.\ Puls$^2$}\\

{\it $^1$University of Hawaii Institute for Astronomy, USA}\\
{\it $^2$Universit\"ats-Sternwarte M\"unchen, Germany}

\begin{abstract}
Recent studies of massive O-type stars present clear evidences of
inhomogeneous and clumped winds. O-type (H-rich) central stars of planetary
nebulae (CSPNs) are in some ways the low mass--low luminosity analogous of
those massive stars. In this contribution, we present preliminary results of
our on-going multi-wavelength (FUV, UV and optical) study of the winds of
Galactic CSPNs. Particular emphasis will be given to the clumping
factors derived by means of optical lines (H$\alpha$ and He{\sc ii}\,4686)
and \textquotedblleft classic\textquotedblright~FUV (and UV) lines.
\end{abstract}
\end{center}

\begin{multicols}{2}

\section{Introduction}
H-rich O-type central stars of planetary nebulae (CSPNs) are 
\textquotedblleft downscaled\textquotedblright~versions
of massive O-type stars, at least with respect to their physical properties 
as derived from spectroscopic studies (see Kudritzki et al. 
\cite{kudritzki2006} and references therein). In the canonical picture of a 
planetary nebula, the fast wind coming from the central star is
extremely important for the structure of the nebula, as well as for the
subsequent evolution of the star itself. So far, the analysis of these winds 
have been based on sophisticated non-LTE models under the consideration of 
homogeneous winds (Kudritzki el al. \cite{kud1997}; Pauldrach et al.
\cite{pauldrach2004}). However,
this assumption seems to be unrealistic, as could be suspected  
from recent studies of the winds of their massive (false) relatives (see 
A. Fullerton, J.C. Bouret, J. Puls or F. Najarro, this volume).
Moreover, independent hints of the presence of inhomogeneous winds have been
presented for the Of-type CSPN NGC\,6543: first, the detection 
of X-ray emission coming from the central star (Chu et al. \cite{chu2001}), 
that could be only explained as the result of the presence of shocks in the
stellar wind (in analogy with massive O-stars) and, secondly, the detection 
of discrete absorption components in FUV/UV profiles (see the 
contribution of R. Prinja in these proceedings).       

Very recently, we have re-analyzed the optical spectra of a sample
of CSPNs by means of non-LTE models atmospheres with inhomogeneous winds 
(Kudritzki et al.\ \cite{kudritzki2006}), computed with {\sc fastwind} 
(Puls et al.\ \cite{puls2005}). It was possible
to estimate wind clumping properties for some of the targets in the sample,
using a novel technique based on the relative strength of H$\alpha$ and
He{\sc ii}\,4686 (see below). In order to check these results, and also to
extend the analysis to other O-type CSPNs, we started a program aimed at the
quantitative analysis of their ultraviolet spectra. In the following, we 
present (some) results of this complementary FUV/UV study on the winds 
of these objects.
 
\section{Spectroscopy of CSPNs}

Due to space constraints, we will not discuss the methodology followed
in the analysis. The reader is referred to any of the many works published in
recent years on quantitative spectroscopy techniques of massive stars. We will 
just provide some detail concerning clumping assumptions.

Regarding the model atmosphere codes, we used {\sc fastwind} for the optical 
analysis, 
and {\sc cmfgen} (Hillier \& Miller\ \cite{hillier1998}) for the FUV/UV
analysis. Clumping is treated presently in both codes under the
{\em micro-clumping} formalism, i.e. {\em small-scale} density
inhomogeneities in the wind redistribute the matter into clumps of
enhanced density, embedded in an almost void (inter-clump) medium.
Clumping is then characterized in the models by the {\em clumping factor} 
$f_\mathrm{cl}$, which represents the overdensity in the clumps with respect
to the smooth medium $\rho_\mathrm{cl}\,=\,f_\mathrm{cl}\,\rho$. Under the
current assumptions, $f_\mathrm{cl}$ corresponds to the inverse of the volume 
filling factor.    

\subsection{Optical analysis}
The analysis of the optical spectra has been presented by Kudritzki et al.\
(\cite{kudritzki2006}). With regard to the distribution of the 
clumping, we assumed a constant clumping factor $f_\mathrm{cl}$ for
velocities larger than twice the characteristic isothermal speed of sound,
and unity for lower velocities. This clumping distribution is basically
equivalent to an exponential law rising (almost) at the base of the wind.

Below Teff$\sim$37--36 kK (depending on the gravity) He~{\sc ii} becomes the 
dominant ionization stage, thus the lines' optical depths present a linear 
dependence with density. On the other hand, neutral hydrogen is a trace ion 
at these Teffs, hence its lines depend on $\rho^2$. Therefore, it would be 
possible to estimate wind  clumping factors by comparing the relative strenthgs 
of H and He{\sc ii} lines with significant wind contribution. In the
optical domain, H$\alpha$ and He{\sc ii}\,4686 are the lines of choice. Using 
this concept, Kudritzki et al.\ (\cite{kudritzki2006}) estimated the 
clumping factors for three CSPNs, ranging from 1 to 50 (see Tab. \ref{urb_tab1}). 
This same technique has been recently applied by Hultzsch et al.\ 
(\cite{hultzsch2007}) to a sample of CSPNs in the Galactic Bulge. 

\begin{table}[H] 
\begin{center} 
\caption{Parameters derived from the analysis of the optical spectra.} 
\label{urb_tab1}
\medskip
\begin{tabular}{ccccc}
\toprule
ID         & Teff  & $\log\,$g & $f_\mathrm{cl}$ & log\,$\dot{M}$  \\
           &  (kK) & (dex)     & & (M$_\odot$\,yr$^{-1}$) \\
\midrule 
IC\,418    & 36 & 3.2 & 50 & -7.43 \\
Hen\,2-108 & 34 & 3.4 &  1 & -7.46 \\
Hen\,2-131 & 32 & 3.2 &  8 & -6.88 \\
\bottomrule
\end{tabular}
\end{center}
\end{table}

\subsection{FUV/UV analysis}
It is a well known fact that there are 
differences regarding the derived parameters when analyzing optical
or UV spectra. These differences are most likely linked to subtle differences
in the preferred codes used in each spectral windows.  
Seeking for consistency, we used our {\sc fastwind} models (the
atmospheric structure, both pseudo-static photosphere and wind) as an input 
for {\sc cmfgen}. In such way, the density distribution used to synthesize
the FUV/UV spectrum is the same that was used for the optical analysis.
Nevertheless, some minor corrections have to be applied to the parameters 
derived from the optical analysis. In particular, and for the domain  
explored here, the S{\sc iv}\,1063--74, C{\sc iii}\,1176 and Si{\sc
iv}\,1394--1402 \AA~ lines present a high sensitivity to Teff. In general, 
these lines require a reduction 
in Teff of $\sim$1--1.5 kK, within the uncertainties of the analysis. 

Concerning clumping, we used the standard implementation in {\sc cmfgen}, an
exponential law. As previously quoted, this is consistent with the constant
clumping form used in the optical analysis, provided that the constant value
is reached soon enough (fast rising). Only in a very limited number of tests
we have tried a distribution with clumping vanishing ($f_\mathrm{cl}$ becoming
unity) in the outer parts of the wind (F. Najarro, this volume). 

We display in Fig. \ref{urb_fig1} line profiles for two different cases,
IC\,418, for which we could infer clumping from the optical analysis, and
IC\,4593, a CSPN too hot to use the H$\alpha$--He{\sc ii}\,4686 method. For
each star, three different lines are presented: the bluest component of the
P{\sc v}\,1118--28 doublet, He{\sc ii}\,1640 and N{\sc iv}\,1719. In the
first case, these three lines change strongly when considering inhomogeneous
wind models. For the second star, the He{\sc ii} line does not change at all:
at this high Teff, He{\sc iii} is the dominant ionization stage, and hence
He{\sc ii} lines behave as H$\alpha$ with respect to wind clumping.
\begin{figure}[H]
\begin{center}
\includegraphics[width=\columnwidth]{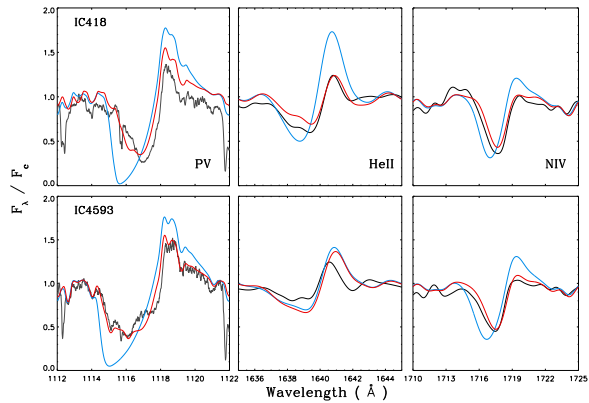}
\caption{Observed (black) spectral lines compared with theoretical
profiles for homogeneous (blue) and clumped (red) winds.\label{urb_fig1}}
\end{center}
\end{figure}

Other lines are also affected by clumping, such as S{\sc vi} 933--45 \AA.
To a lesser degree, and most likely as a consequence of an \textquotedblleft
indirect\textquotedblright~effect (due to the reduced $\dot{M}$),
photospheric lines (Fe{\sc v} and O{\sc iv}) display variations in the sense 
that the fits are improved when (wind) clumping is considered.    

All the CSPNs in our sample for which
FUV data (either FUSE, COPERNICUS or TUES) are available present features
that can be associated with the O{\sc vi}\,1032--38 \AA~lines. These
{\em super-ionization} features (as well as N{\sc v}\,1238--42 \AA) are related 
to X-rays, usually explained as produced by shocks in the winds. We have
tried to model these features in some (few) cases, to investigate the sensitivity
of the primary clumping diagnostics to the presence of X-rays. While we
did not manage to produce completely satisfactory fits, these tests have
shown that all the clumping indicators are insensitive to the presence of
X-rays (note that present implementations of X-rays in model atmosphere
codes are very crude). 

\section{Discussion}

While we have results presently for a handful of objects, there are a number of
conclusions that can be drawn from this preliminary work. First, it seems to be 
possible
to achieve good fits to FUV/UV spectra with the parameters derived from the 
optical analysis (see Fig. \ref{urb_fig3}). The ability of reproducing 
simultaneously ultraviolet and optical
ranges increases our confidence on that the physics considered in the models
is a fair representation of the true one (i.e. we are not missing any
important contribution). Secondly, FUV/UV clumping sensitive lines support 
$f_\mathrm{cl}$ values derived using H$\alpha$--He{\sc ii}\,4686 for the
coolest objects. There is not an apparent reason why this method should not
work also for massive O-stars in the appropriated Teff--log\,g domain.
Thirdly, and most important, CSPNs with very similar fundamental parameters 
have substantial differences in their clumping properties. 
To the best of our knowledge, this has not been 
found (yet) in the case of massive O-stars. Should this be confirmed, it would 
have tremendous implications for (theoretical) predictions of CSPNs mass-loss 
rates.

\begin{figure}[H]
\begin{center}
\includegraphics
  [width=\columnwidth]{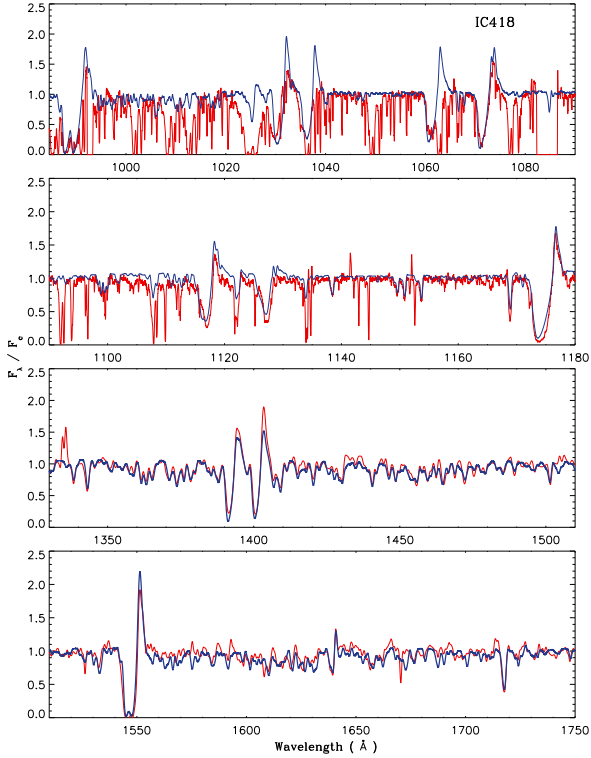}
\caption{IC\,418 FUSE and IUE data (red). A model including clumping 
($f_\mathrm{cl}\,=\,50$) and X-rays 
($\log\,\mathrm{L}_\mathrm{x}\,/\,\mathrm{L}_\odot\,=\,-4.27$) is shown 
in blue. Note the heavy
contamination of FUSE ranges by the medium surrounding the star. \label{urb_fig3}}
\end{center}
\end{figure}

There are a number of open issues related to the comparison of the derived
properties with the corresponding expected theoretical values, from the
point of view of post-AGB evolution as well as from the radiatively driven
wind (RDW) theory. First, for some of the objects we still derive uncomfortably 
high spectroscopic masses. Since these objects are evolving directly to the WD phase,
CSPN masses above $\sim$0.8--0.9\,M$_\odot$ (as derived for IC\,418, Tc\,1
and NGC\,2392) seem unrealistic (WD mass distribution peaks around
$\sim$0.6 M$_\odot$). Secondly, the ratios of the measured wind terminal
velocities to the derived escape velocities are in general higher than the
values expected from the RDW theory. Increasing the masses will bring these
ratios closer to the expected values, but this would badly affect the
comparison with post-AGB evolutionary models, increasing the number of objects
with extremely high spectroscopic mass. 

At present, it is not clear where the solution to these two problems resides.
Is there any important ingredient missing in our model atmospheres? As
previously quoted, our ability to reproduce a wide spectral range
(FUV/UV/optical) seems to argue against this possibility, although it
cannot be completely ruled out. On the other hand, wind hydrodynamics
are based on the assumption of smooth winds. Theoretical 
predictions have to be checked with the inclusion of clumped
winds (deKoter and Kriticka, this volume).


\end{multicols}

\end{contribution}


\end{document}